\documentclass[10pt,aps,prb,twocolumn,superscriptaddress]{revtex4-2}

\usepackage[T1]{fontenc}
\usepackage{amssymb,amsmath}
\usepackage{graphicx}
\usepackage{color}
\usepackage{hyperref}
\usepackage{listings}
\usepackage{siunitx}
\usepackage{gensymb}
\usepackage{textcomp}
\usepackage{xspace}
\usepackage{lipsum}
\usepackage{upgreek}

\usepackage{subcaption}
\newcommand{\micron}{{$\mu$}m\xspace}
\begin{document}


\title{Changes in electric-field noise due to thermal transformation of a surface ion trap}

\author{Maya Berlin-Udi }
\affiliation{Department of Physics, University of California, Berkeley, California 94720, USA}
\affiliation{Challenge Institute for Quantum Computation, University of California, Berkeley, CA  94720}
\author{Clemens Matthiesen}
\affiliation{Department of Physics, University of California, Berkeley, California 94720, USA}
\author{P. N. Thomas Lloyd}
\thanks{Current address: Department of Physics, University of California, Santa Barbara, California 93106, USA}
\affiliation{Department of Physics, University of California, Berkeley, California 94720, USA}

\author{Alberto M. Alonso}
\affiliation{Department of Physics, University of California, Berkeley, California 94720, USA}
\affiliation{Challenge Institute for Quantum Computation, University of California, Berkeley, CA  94720}
\author{Crystal Noel}
\thanks{Current address: Duke Quantum Center, Department of Electrical and Computer Engineering, Duke University, Durham, NC 27708, USA}
\affiliation{Department of Physics, University of California, Berkeley, California 94720, USA}
\author{Benjamin Saarel}
\affiliation{Department of Physics, University of California, Berkeley, California 94720, USA}
\affiliation{Challenge Institute for Quantum Computation, University of California, Berkeley, CA  94720}
\author{Christine A. Orme}
\affiliation{Lawrence Livermore National Laboratory, Livermore, CA, 94551, USA}
\author{Chang-Eun Kim}
\affiliation{Lawrence Livermore National Laboratory, Livermore, CA, 94551, USA}
\author{Art J. Nelson}
\affiliation{Lawrence Livermore National Laboratory, Livermore, CA, 94551, USA}
\author{Keith G. Ray}
\affiliation{Lawrence Livermore National Laboratory, Livermore, CA, 94551, USA}
\author{Vincenzo Lordi}
\affiliation{Lawrence Livermore National Laboratory, Livermore, CA, 94551, USA}
\author{Hartmut H\"affner}
\affiliation{Department of Physics, University of California, Berkeley, California 94720, USA}
\affiliation{Challenge Institute for Quantum Computation, University of California, Berkeley, CA  94720}
\email{hhaeffner@berkeley.edu}

\date{\today}

\begin{abstract}

We aim to illuminate how the microscopic properties of a metal surface map to its electric-field noise characteristics. In our system, prolonged heat treatments of a metal film can induce a rise in the magnitude of the electric-field noise generated by the surface of that film. We refer to this heat-induced rise in noise magnitude as a thermal transformation. The underlying physics of this thermal transformation process is explored through a series of heating, milling, and electron treatments performed on a single surface ion trap. Between these treatments, $^{40}$Ca$^+$ ions trapped 70~$\mu$m above the surface of the metal are used as detectors to monitor the electric-field noise at frequencies close to 1~MHz. An Auger spectrometer is used to track changes in the composition of the contaminated metal surface. With these tools we investigate contaminant deposition, chemical reactions, and atomic restructuring as possible drivers of thermal transformations.

\end{abstract}

\maketitle

\section{Introduction}

Surface electric-field noise describes the fluctuation of an electric field due to charge movement in the surface of a material. Noise from metal surfaces is of particular interest to the ion trapping community, as metal electrodes, necessary to define the electric fields in an ion trap, constitute the nearest surfaces to the ions. Such ions are trapped in harmonic potentials, and electric-field noise resonant with the motional modes of the trap can cause ion heating and motional decoherence. These processes are detrimental to applications of trapped ions including quantum information processing and tests of fundamental physics~\cite{Brownnutt2015, Brown2020-materials}.

While it is generally understood that dynamic processes such as defect hopping and contaminant diffusion take place at metal surfaces, it is not known which microscopic processes specifically dominate charge dynamics and lead to surface electric-field noise. Advancing our understanding of surface charge dynamics could open new avenues for ion trap engineering. It could also benefit other technologies which suffer from charge noise or surface noise of various kinds, both within the field of quantum information processing, and beyond. This includes solid-state qubits in diamond~\cite{Kim2015, Myers2017Double-QuantumCenters}, superconducting qubits~\cite{Paladino2014Information, Christensen2019_chargenoise}, single-electron transistors~\cite{Kafanov2008}, Casimir force detection~\cite{Garrett2015TheMicroscopy}, and precise measurements of gravitational fields~\cite{Antonucci2012InteractionMass, Armano2017}.

Measurements of electric-field noise in ion traps have yielded a wide variety of results between different research groups and different traps across a range of studied parameters, such as the noise magnitudes, frequency and distance dependencies, and the response to surface treatments~\cite{Brownnutt2015}. Such variations indicate that no single physical process is responsible for all surface noise. Most likely, in any given system there are multiple competing mechanisms at play~\cite{Sedlacek2018-multi-mechanisms}. To gain a general understanding of electric-field noise from metal surfaces, we must first disentangle these mechanisms through in-depth studies of individual surfaces.

Several different types of surface treatments have been performed on ion traps in various efforts to understand and reduce noise. Laser ablation reduced electric-field noise by about 50\% in one ion trap \cite{Allcock2011-heating-reduction}. Ion milling reduced noise by up to two orders of magnitude in several investigations~\cite{Daniilidis2014-ion-milling, Hite2012, McKay2014b, Sedlacek2018-multi-mechanisms}, and oxygen plasma reduced noise by a factor of four according to one report~\cite{McConnell2015-plasma-cleaning}. In all cases, these treatments removed contamination from the surface of the trap electrodes. These results suggest that surface contaminants may play an important role in generating electric-field noise in ion traps. 

However, the relationship between contamination and electric-field noise is not straightforward.
Daniilidis {\em et al.}~\cite{Daniilidis2014-ion-milling} found that a trap recontaminated after milling still maintained a low noise magnitude. Sedlacek {\em et al.}~\cite{Sedlacek2018-multi-mechanisms} observed that for some materials, ion milling actually increased noise at cryogenic temperatures, and Kim {\em et al.}~\cite{Kim2017Theory-and-Experiments} found that noise first rose and then fell in response to a series of small-dose ion milling treatments.

Given that noise magnitudes have not been found to be directly proportional to contaminant levels on ion trap surfaces, it is likely that contaminant removal is not solely responsible for the observed treatment-induced noise reductions. Here we aim to determine how ion milling and other surface treatments can affect surface noise levels at a microscopic level. By understanding the mechanisms behind noise reduction and promotion, we can determine which surface properties correspond to high and low noise levels. We anticipate that this information will help in the design of low-noise quantum devices.

We use trapped $^{40}$Ca$^+$ ions as detectors to study the noise generated by the surface of an aluminum-copper ion trap. We measure noise at multiple temperatures and frequencies, and then observe how the noise responds to changes in the properties of the trap surface. The surface properties are altered using treatments including prolonged heating, argon ion milling, and electron bombardment. The trap surface composition is tracked via \textit{in situ} Auger spectroscopy. As shown in a previous publication, the measured noise characteristics for our trap are consistent with noise produced by an ensemble of thermally activated fluctuators~\cite{Noel2019-TAF}, so our data are discussed in this context. 

We have observed that argon ion milling reduces noise at frequencies near 1-MHz, while altering the composition of the trap surface. We have further observed that prolonged heat treatments can increase the room-temperature noise magnitude without altering the measured surface composition. The latter provides us with an opportunity to study how noise is correlated to a surface property other than the contaminant level or temperature, which have both been previously studied. 

In this manuscript, we will explore the underlying physics of the thermal transformations observed in our system. A thermal transformation is a sustained change in the noise spectrum (over a defined temperature and frequency range), caused by the thermally driven modification of the physical properties of the noise source. Such modifications may include processes such as the addition, removal, or restructuring of material. A thermal transformation is distinct from a thermal noise activation. When noise is thermally activated, the noise spectrum changes in a predictable and reversible fashion determined by the characteristic activation energy. Thermally activated noise thus has a repeatable temperature dependence, and might stem from noise sources such as thermally activated fluctuators \cite{Noel2019-TAF} or fluctuating dipoles \cite{Ray2019}. In contrast, noise that is being thermally transformed does not have a repeatable temperature dependence, i.e. the characteristics of the noise source are permanently altered by the thermal process.

Labaziewicz {\em et al.}~\cite{Labaziewicz2008} found that high-temperature, \textit{ex situ} annealing could reduce noise measured in ion traps at cryogenic temperatures. To the best of our knowledge, our experiments are the first to explore the impact of annealing on noise produced by ion traps at or above room temperature. 

The manuscript is organized as follows. In Sec.~\ref{sec:methods}, we discuss the methods we use to treat and characterize the ion trap surface. In Sec.~\ref{sec:data} we present results from a wide range of measurements taken throughout our study of thermal noise transformations. These include measurements of ion heating rates as a function of temperature, which can be directly translated to an electric-field noise temperature scaling. Additional measurements include Auger spectra, which provide information on the atomic and chemical composition of the surface, and \textit{ex situ} AFM data which can be used to characterize the surface morphology. We also report how ion heating rates change in response to heat treatments, electron bombardment, and argon ion milling. In Sec.~\ref{sec:analysis} we analyze and interpret our measurement results, and explore contaminant deposition, chemical reactions and atomic restructuring as possible drivers of thermal noise transformations.

\section{Methods} \label{sec:methods}

We perform a series of \textit{in situ} surface treatments on a single rf quadrupole surface ion trap without breaking vacuum. The base pressure of the vacuum chamber is about $1\times 10^{-10}$~Torr. The surface trap under investigation is composed of an insulating fused silica substrate and a conductive metal surface. Trenches are etched into the fused silica to define (and provide electrical isolation between) rf and dc electrodes. These electrodes are biased or driven in order to generate the confining electric fields necessary for trapping ions. 

Six layers of metal were deposited onto the trap surface via electron-beam evaporation. The first layer is 15~nm thick and composed of titanium to act as a sticking agent, followed by 500~nm of aluminum and 30~nm of copper. After these three layers were deposited, the trap was briefly exposed to atmosphere as the trap was rotated, and then an additional 15~nm of titanium, 500~nm of aluminum and 30~nm of copper were deposited. Subsequent heating in the experiment vacuum chamber caused the metal layers to intermix. The trap was stored in atmosphere both before and after baking for a cumulative exposure time of 30 weeks, causing the trap surface to oxidize and accumulate contaminants.

This trap is designed to hold ions in a harmonic potential 70~\micron above the metal surface. The motional modes of a trapped ion have well-defined frequencies in the MHz range.  When exposed to electric-field noise at these frequencies, the ion motion heats up. Measuring the heating rate of the motion allows us to use the ion as an ultra-sensitive narrow-band electric-field noise detector. We measure the motional heating rate of a single trapped $^{40}$Ca$^+$ ion by tracking the decay of carrier Rabi oscillations \cite{roosthesis}. These ions are Doppler cooled to energies of approximately 10 vibrational quanta before measurement. The electric-field noise power spectral density $S_E(\omega)$ is directly proportional to the ion motional heating rate $\dot{\bar{n}}$ according to \cite{Brownnutt2015}: 

\begin{equation}
S_E(\omega) = \frac{4m\hbar \omega}{q^2}\dot{\bar{n}}(\omega) ,
\label{eqn:hr}
\end{equation}

where $m$ and $q$ are the mass and the charge of the ion, $\hbar$ is the reduced Planck constant, and $\omega$ is the secular trap frequency. $S_E(\omega)$ is a measure of the amplitude of the electric-field noise at frequency $\omega$.

\begin{figure}[ht]
	\centering
	\includegraphics{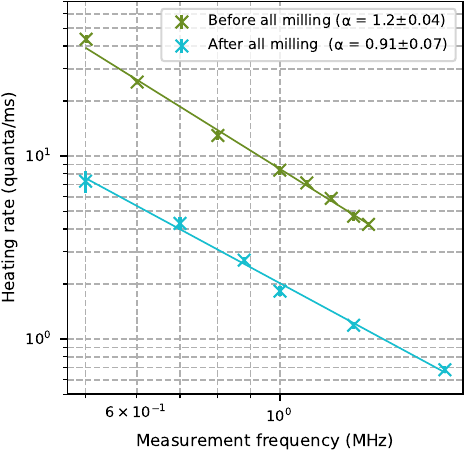}
	\caption{Ion heating rates measured as a function of frequency at room temperature. The heating rate scales inversely with frequency. The frequency scaling exponent $\alpha$ is close to one, both before and after surface treatment. The data labeled \textit{Before} and \textit{After} were measured before \textsc{heat 1} and \textsc{heat 10} respectively.}\label{fig:freqscalings}
\end{figure}

By altering the ion confinement potential, we can measure noise at frequencies anywhere between $2 \pi \times 0.5$~MHz and $2 \pi \times 2$~MHz. Two sets of measurements are shown in Fig.~\ref{fig:freqscalings}. We find that the ion heating rates follow a power-law dependence as

\begin{equation}
\dot{\bar{n}} \propto 1/f^{\alpha+1} ,
\label{eqn:freq_scale}
\end{equation}

where $f$ is the noise frequency and $\alpha$ is the frequency scaling exponent. According to Eq.~\eqref{eqn:hr}, the surface electric-field noise in our system then scales as $1/f^{\alpha}$. We measured $\alpha$ to be close to 1 both before and after surface treatments. This suggests that these surface treatments may not fundamentally change the nature of the dominant noise mechanisms.

The trap is clamped to a resistive button heater, which enables us to measure how the surface noise depends on the temperature of the ion trap chip. We monitor the trap temperature using a thermal imaging camera. This experimental setup is described in Ref.~\cite{Noel2019-TAF}. We are able to measure ion heating rates when the chip temperature is between room temperature and 600~K. At higher temperatures, the ion lifetime becomes too short to perform measurements. 

\begin{figure}[ht]
	\centering
	\includegraphics{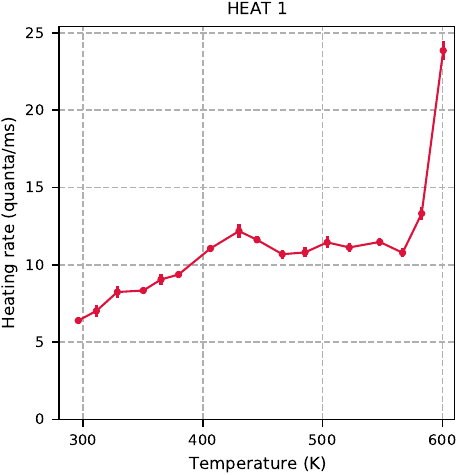}
	\caption{ Ion heating rates measured as a function of ion trap substrate temperature. Measurements were performed at trap frequency $2 \pi \times 1.3$~MHz.}\label{fig:full_range_temp_scaling}
\end{figure}

Ion heating rates measured as a function of the temperature of our ion trap chip are plotted in Fig.~\ref{fig:full_range_temp_scaling}. At approximately 575~K, heating rates begin to rise rapidly. This high-temperature behavior endured qualitatively throughout all surface treatments, and was uncorrelated with the presence of thermal transformations. This may be evidence that a new noise mechanism becomes dominant above 575~K in our trap.  In this manuscript we will focus only on the mechanisms dominant at lower temperatures and report heating rates measured up to 550~K in the following~\footnote{See the Berlin-Udi dissertation \cite{Berlin-Udi2020TreatmentsSurface} for more discussion of heating rates measured in this trap above 550~K. This dissertation also contains additional information about the trap geometry, measurement methods, analysis techniques, and the experimental setup.}.

We will focus on ten heat treatments that we performed on a single surface ion trap. A heat treatment involves the heating of the substrate to temperatures greater than 500~K for a period of days. The parameters of these heat treatments are summarized in Tab. \ref{ch:heat_tab:heat}, and plotted in detail in App.~\ref{ap:history}. Fig.~\ref{fig:full_range_temp_scaling} shows data recorded at the end of treatment \textsc{heat 1}. Each heat treatment was immediately preceded by argon ion milling, with the exception of \textsc{heat 1}, which was performed before any milling took place.

\begin{center}
\begin{table}[h!]
  \begin{center}
    \caption{Summary of the temperatures and timings of ten heat treatments performed on our ion trap. The \textit{cumulative material removed} refers to estimates (see text) of the total thickness of material that was removed from the surface of the trap via argon ion milling between the trap's fabrication and the start of each heat treatment. Additional details of the heat treatment procedures are plotted in App. \ref{ap:history}.}
    \label{ch:heat_tab:heat}
    \begin{tabular}{| l | S | S | S | S | S}
      \hline
       \textbf{Treatment} & \textbf{Time}   & \textbf{Time}   & \textbf{Cumulative}    & \textbf{Most}\\  
                          & \textbf{over}   & \textbf{over}   & \textbf{material} & \textbf{recent}  \\  
	  		              & \textbf{550~K}  & \textbf{575~K}  & \textbf{removed}  & \textbf{mill dose}\\
                          & \textbf{(days)} & \textbf{(days)} & \textbf{(nm)}     & \textbf{(J/cm$^2$)} \\ 
      \hline
	  \textsc{heat 1}	 &   2.5    &  2   &  0     & 0   \\
  	  \textsc{heat 2}    &   3      &  3   &  0.13  & 0.1 \\
  	  \textsc{heat 3}    &   2      &  0.3 &  0.26  & 0.1 \\ 
  	  \textsc{heat 4}    &   1.5    &  1   &  0.86  & 0.5 \\
  	  \textsc{heat 5}    &   5      &  3   &  6.3   & 2.9 \\
  	  \textsc{heat 6}    &   4      & 0.2  &  17.7  & 7.5 \\ 
  	  \textsc{heat 7}    &   7      &  4   &  34    & 8.8 \\
  	  \textsc{heat 8}    &   4.5    &  3.5 &  36    & 0.6 \\
      \textsc{heat 9}    &   3      &  3   &  48    & 4.9 \\
      \textsc{heat 10}   &  20      &  9   &  77    & 5.2 \\
      \hline
      \end{tabular}
      \end{center}
\end{table}
\end{center}

We performed argon ion milling with ion energies between 200 and 500~eV. The argon pressure in the chamber during these treatments was approximately $10^{-5}$~mbar, and ion currents ranged from 5 to 10 \micro A/cm$^2$. The ion beam was angled normal to the trap surface during all treatments, except in the final milling step performed before \textsc{heat 10}. During this final milling step, the beam was positioned at a 45$\degree$ angle with respect to the trap surface, as well as to the trap measurement axis. Each milling treatment was performed for a different period of time. The details of each milling treatment can be found in App.~\ref{ap:history}. We used the SRIM software \cite{Ziegler2010} to determine the amount of material removed in each milling step, taking into account ion energy, ion current, milling duration, and surface composition as measured with Auger spectroscopy. The estimated cumulative material removed by the start of each heat treatment is shown in Tab.  \ref{ch:heat_tab:heat}. The cumulative milling energy for all argon ion treatments was 42~J/cm$^2$.

\begin{figure*}[ht]
	\centering
	\begin{subfigure}[t]{.23\textwidth}
		{\includegraphics{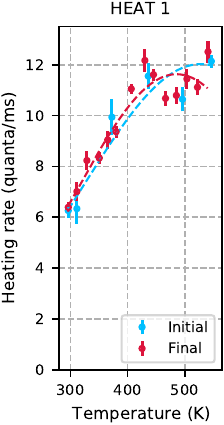}}
	\end{subfigure}%
	\hskip 2ex
	\begin{subfigure}[t]{.168\textwidth}
		{\includegraphics{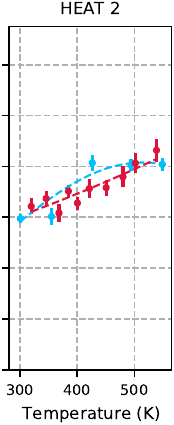}}
	\end{subfigure}%
	\hskip 2ex
	\begin{subfigure}[t]{.168\textwidth}
		{\includegraphics{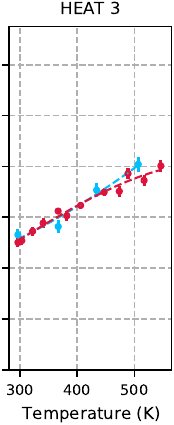}}
	\end{subfigure}%
	\hskip 2ex
	\begin{subfigure}[t]{.168\textwidth}
		{\includegraphics{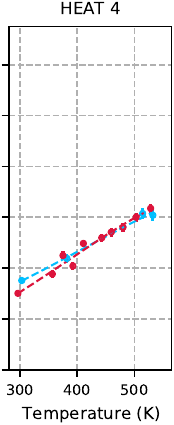}}
	\end{subfigure}%
	\hskip 2ex
	\begin{subfigure}[t]{.168\textwidth}
		{\includegraphics{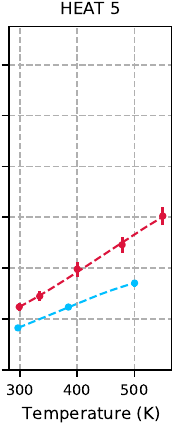}}
	\end{subfigure}%
	\vskip 2ex
	\begin{subfigure}[t]{.22\textwidth}
		{\includegraphics{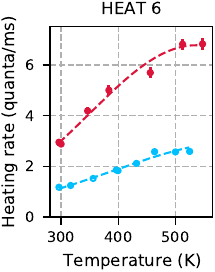}}
	\end{subfigure}%
	\hskip 2ex
	\begin{subfigure}[t]{.17\textwidth}
		{\includegraphics{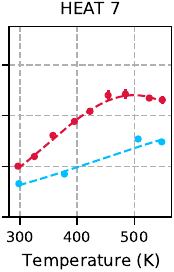}}
	\end{subfigure}%
	\hskip 2ex
	\begin{subfigure}[t]{.17\textwidth}
		{\includegraphics{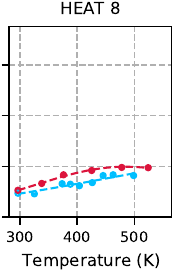}}
	\end{subfigure}%
	\hskip 2ex
	\begin{subfigure}[t]{.17\textwidth}
		{\includegraphics{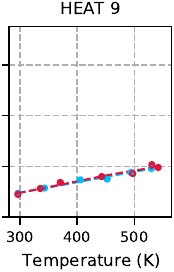}}
	\end{subfigure}%
	\hskip 2ex
	\begin{subfigure}[t]{.17\textwidth}
		{\includegraphics{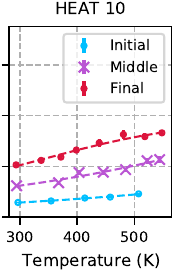}}
	\end{subfigure}%

	\caption{Ion heating rates measured as a function of substrate temperature during ten heat treatments. The \textit{initial} heating rates were measured as the temperature of the substrate was raised at the start of each treatment. The \textit{final} heating rates were measured after the substrate was held at temperatures greater than 500~K for multiple days.
	Thermal transformations are observed in \textsc{heat 5, 6 ,7, 10} and to a lesser degree \textsc{heat 8}. The trap was milled with argon ions between heat treatments. Information on milling and heat treatment parameters can be found in App. \ref{ap:history} and in Tab. \ref{ch:heat_tab:heat}. The  \textsc{heat} 1, 3, 4, 6, 7, 8, and 9 measurements plotted here were performed at trap frequency $2 \pi \times 1.3$~MHz. The \textsc{heat} 2, 5, and 10 measurements plotted here were performed at trap frequency $2 \pi \times 0.88$~MHz and scaled to $2 \pi \times 1.3$~MHz assuming $\alpha = 1$ for the purposes of this figure. } \label{fig:recrystal_all_tempscalings}
\end{figure*}

An Auger spectrometer is installed \textit{in situ}, so that heat treatments, argon ion milling, ion heating rate measurements and surface composition measurements can all be performed without breaking vacuum. We use an OCI Vacuum Microengineering Inc. Auger spectrometer with a retarding field analyzer, and a lock-in amplifier that allows us to record the differential Auger spectrum. This method is sensitive to the top 3 to 10 monolayers of a substrate, depending on its material composition. To determine the fractional composition of the trap surface, we match the measured peaks to spectra of known elements, then measure the peak-to-peak magnitude for each element, and scale by a sensitivity factor provided by the manufacturer of the spectrometer.

Our Auger spectroscopy involves bombarding the trap surface with 2~keV electrons, and then measuring the amplitude of Auger electrons emitted from the trap surface. At the center of the electron beam, the current density is between 3 and 12~mA/cm$^2$. This incident electron beam can also be used as a surface treatment in its own right. When performing an electron treatment, the electron beam is aligned such that it bombards the trap surface directly below where the calcium ions can be trapped. When we want to measure the surface composition without altering the the surface at the trapping position,  we record an Auger spectrum with the electron beam positioned away from the ion-trapping area.

\section{Results} \label{sec:data}

We measured the heating rate as a function of substrate temperature during ten heat treatment experiments. The measurement results are plotted in Fig.~\ref{fig:recrystal_all_tempscalings}. The trap was milled with argon ions before each heat treatment, and this milling caused ion heating rates to steadily decrease in magnitude. In some temperature-dependence measurements, the heating rate increased monotonically with temperature. In other experiments the temperature dependence saturated at high temperatures, or, in the case of \textsc{heat 7} Final, decreased slightly at high temperatures.

In some of these experiments, the temperature dependence of the ion heating rate did not change as a result of the heat treatment. In other experiments, the substrate's prolonged exposure to high temperatures caused thermal transformations to take place, and the ion heating rates increased across the measured temperature range. Thermal transformations took place during treatments \textsc{heat} 5, 6, 7, 10 and, to a lesser extent, \textsc{heat 8}. Details of the timings and temperatures of heat treatments are presented in App.~\ref{ap:history}, and a summary can be found in Table \ref{ch:heat_tab:heat}.

Following this overview of the heating rate data, we present results from a range of surface treatments with a focus on how they relate to the thermal transformation process.

\begin{figure}[ht]
	\centering
	\begin{subfigure}[t]{.5\textwidth}
		\caption{}
	    \vspace{-.3 cm}
	    \label{fig:hr_v_time_recrystal_m7}
		\includegraphics{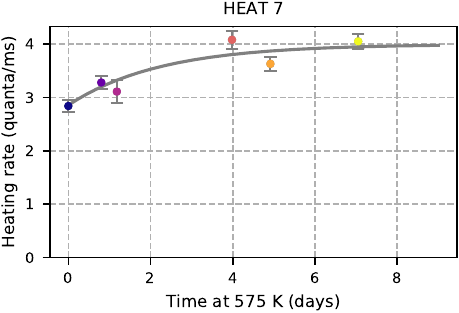}
	\end{subfigure}%
	\hskip 2ex
	\begin{subfigure}[t]{.5\textwidth}
	    \caption{}
	    \vspace{-.3 cm}
	    \label{fig:temp_v_time_recrystal_m7}
		\includegraphics{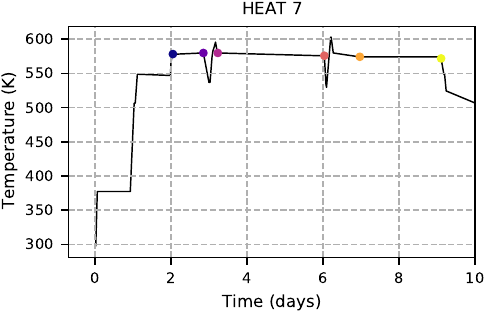}
	\end{subfigure}%

	\caption{Ion heating rates measured during treatment \textsc{heat 7} while the substrate temperature was held at 575~K show a saturating noise transformation with a half life of $1.6\pm0.8$~days. (a) The time dependence of ion heating rates, fit to Eq. \ref{eq:heat_treat}. (b) Timing and temperatures of treatment \textsc{heat 7}. To determine the context in which each measurement was taken, match the colors of the data plotted in (a) to the colors of the data in (b). These heating rates were measured at $2 \pi \times 1.3$~MHz. }\label{fig:m7_v_time_recrystal}
\end{figure}

\paragraph*{Thermal transformation dynamics:}
To quantify the timescale of the thermal transformation process, we raised the substrate temperature relatively swiftly to 575~K at the start of \textsc{heat 7}, and repeatedly measured the ion heating rates at 575~K for a full week. Heating rates rose steadily for the first few days, and then the noise transformation saturated. The parameters of this heat treatment, and the corresponding ion-heating-rate measurement results, are presented in Fig.~\ref{fig:m7_v_time_recrystal}. Subsequent temperature cycling caused the heating rate at 550~K to rise by an additional 12\%.

The constant-temperature heating rate measurements from  \textsc{heat 7}, as plotted in Fig.~\ref{fig:m7_v_time_recrystal}, are fit to a simple exponential saturation function:

\begin{equation} \label{eq:heat_treat}
	\Gamma = \Gamma_0 + A(1-e^{-kt})
\end{equation}
where $\Gamma$ is the ion heating rate at 575~K, $t$ is the time spent at 575~K, and $\Gamma_0$, amplitude $A$ and rate constant $k$ are fit parameters. The half life $\lambda$ of this transformation can be calculated directly from the rate constant $k$ as $\lambda = $~ln(2)/$k$, yielding a half life of $1.6 \pm 0.8$~days for the noise transformation at 575~K.

\begin{figure}[ht]
	\centering
	\begin{subfigure}[t]{.5\textwidth}
	    \caption{}
	    \vspace{-.3 cm}
		\label{fig:hr_v_time_recrystal_m12}
		\includegraphics{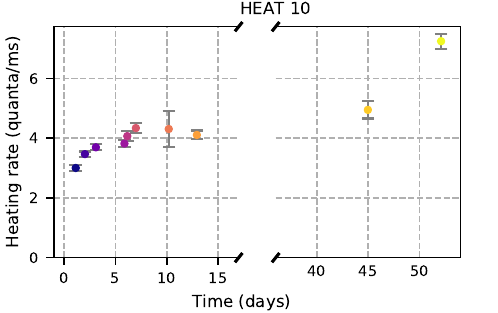}
	\end{subfigure}%
	\hskip 2ex
	\begin{subfigure}[t]{.5\textwidth}
	    \caption{}
	    \vspace{-.3 cm}
	    \label{fig:temp_v_time_recrystal_m12}
	    \includegraphics{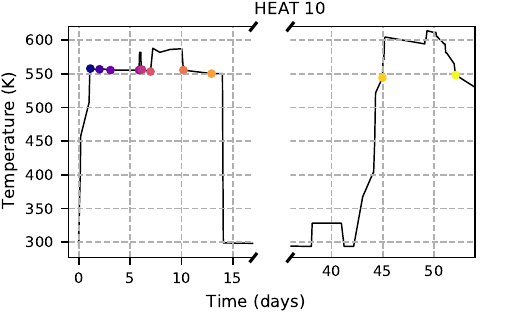}
	\end{subfigure}%
	
	\caption{Ion heating rates were repeatedly measured at 550~K during treatment \textsc{heat 10}. (a) The time dependence of ion heating rates. (b) Timing and temperatures of treatment \textsc{heat 10}. To determine the context in which each measurement was taken, match the colors of the data plotted in (a) to the colors of the data in (b). These heating rates were measured at $2 \pi \times 0.88$~MHz.}\label{fig:m12_v_time_recrystal}
\end{figure}

During treatment \textsc{heat 10}, the trap temperature was held between 555 and 580~K for twelve days, as shown in Fig.~\ref{fig:m12_v_time_recrystal}. During this time, we measured the ion heating rate repeatedly at 555~K. The heating rate initially appeared to saturate. However, after subsequent temperature cycling the heating rate had increased. Raising the trap temperature above 600~K for a period of 3.5 days, we found the ion heating rate rose by an additional 50\%.

\begin{figure}[ht!]

	\begin{subfigure}[t]{.5\textwidth}
	    \caption{}
		\vspace{-.3 cm}
	    \label{ch:heat_fig:big_mill}
		\includegraphics{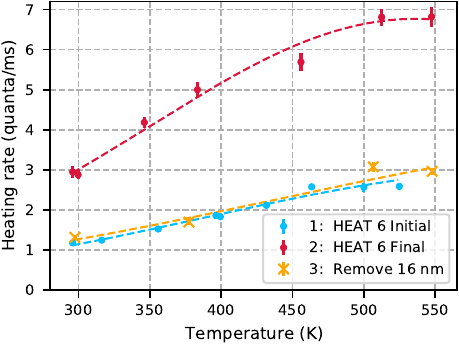}
	\end{subfigure}	\\
	\begin{subfigure}[t]{.5\textwidth}
		\caption{}
		\vspace{-.3 cm}
		\label{ch:heat_fig:little_mill}
		\includegraphics{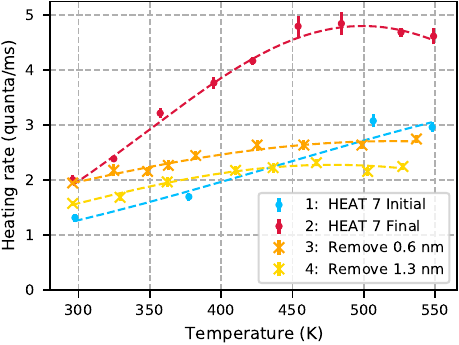}
	\end{subfigure}%
	
	\caption{(a) Treatment \textsc{heat 6} drove a thermal transformation which raised the ion heating rates. The subsequent ion milling removed 16~nm of material and brought the heating rates back down. (b) Treatment \textsc{heat 7} drove a thermal transformation which raised ion heating rates. The subsequent small milling doses removed a total of 2~nm of material, and brought the heating rates down nearly to pre-transformation levels. These heating rates were measured at $2 \pi \times 1.3$ MHz. }\label{ch:heatfig:mill_two_ways}
\end{figure}

\paragraph*{Ion milling of thermally transformed surfaces:}
The surface of the trap was milled with argon ions between each heat treatment. After \textsc{heat 6}, during which a significant thermal transformation took place, 16~nm of material was milled away from the trap surface. As shown in Fig.~\ref{ch:heat_fig:big_mill}, this milling treatment brought the ion heating rates down to pre-transformation levels. 

We performed a series of small-dose milling treatments after the \textsc{heat 7} thermal transformation, removing 0.6~nm and then 1.3~nm of material from the trap surface. The incident argon ions had an energy of 200~eV, and we estimate that they penetrated about 2~nm into the metal \cite{Ziegler2010}. The temperature dependence of the ion heating rate was measured after each of these treatments. As shown in Fig.~\ref{ch:heat_fig:little_mill}, these small milling treatments had a significant impact on the ion heating rate. Although only a few nanometers of material was affected by milling, the heating rates were brought down nearly to pre-transformation levels.

\begin{figure*}[ht!]
	\begin{subfigure}[t]{.45\textwidth}
		\caption{}
		\label{fig:electron3}
		\includegraphics{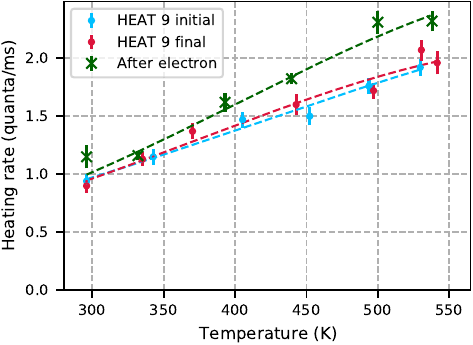}
	\end{subfigure}%
	\hskip 2ex
	\begin{subfigure}[t]{.45\textwidth}
		\caption{}
		\label{fig:electron4}	
		\includegraphics{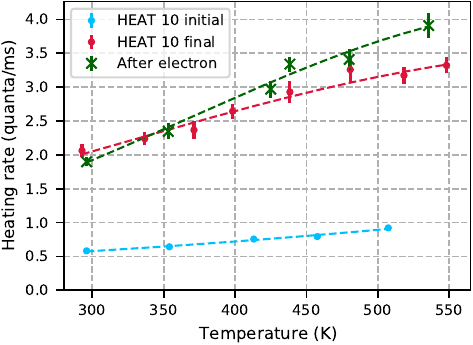}
	\end{subfigure}%
	\caption{Temperature dependence of ion heating rates measured before and after the trap surface was bombarded by a 2~keV electron beam for 3.2 hours. (a) Electron treatment of 180 kJ/cm$^2$ performed after \textsc{heat 9}, a heat treatment in which the noise did not thermally transform. These heating rates were measured at $2 \pi \times 1.3$ MHz. (b) Electron treatment of 280 kJ/cm$^2$ performed immediately after \textsc{heat 10}, when the noise underwent a significant thermal transformation. These heating rates were were measured at $2 \pi \times 0.88$ MHz and scaled to a frequency of $2 \pi \times 1.3$ MHz assuming $\alpha$ = 1 for the purposes of this figure.
	} \label{fig:electron_plots}
\end{figure*}

\paragraph*{Electron bombardment:}
Electron treatments were performed on the trap surface immediately after \textsc{heat 9}, when the trap was heated but not transformed, and immediately after \textsc{heat 10}, when the trap underwent a thermal transformation. We measured the temperature dependence of the ion heating rates before and after these electron treatments, and the results are plotted in Fig.~\ref{fig:electron_plots}. The change in the heating-rate temperature dependence due to electron treatments was comparable in both cases, with ion heating rates slightly increasing at high temperatures as compared to the pre-electron-treatment measurements.

\begin{figure*}[ht!]
	\centering
	\includegraphics{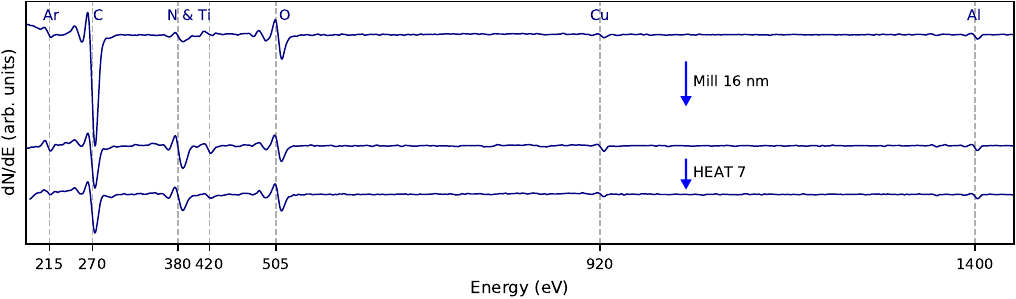}
	\caption{Auger spectra measured before and after argon ion milling and treatment \textsc{heat~7} show features consistent with the presence of argon, carbon, nitrogen, titanium, oxygen, copper, and aluminum. This milling treatment had a significantly greater impact on the surface composition than the subsequent heat treatment. There is no evidence of new elements emerging during the heat treatment. dN/dE is the derivative of the magnitude of Auger electron emission with respect to Auger electron energy. }\label{fig:auger_spectra_recrystal}
\end{figure*}

\paragraph*{Auger spectroscopy:}
Auger spectra were measured at various times throughout our experiments in an effort to characterize the composition of the trap surface. Data from three of these measurements are plotted in Fig.~\ref{fig:auger_spectra_recrystal}. The peaks visible in these spectra indicate the presence of aluminum, copper, titanium, oxygen, carbon, nitrogen and argon. Comparing the spectra measured before and after the \textsc{heat 7} thermal transformation (middle and bottom curves), we deduce that no significant quantities of new elements were introduced to the trap surface during the heat treatment, with the caveat that Auger spectroscopy is not sensitive to hydrogen. In addition, the relative peak heights of different elements are nearly identical before and after heating. This is the case for all spectra measured before and after heat treatments, as shown in App.~\ref{ap:auger_raw}. In contrast, milling the surface had a significant impact on its composition (top and middle curves in Fig.~\ref{fig:auger_spectra_recrystal}).

\begin{figure}[ht!]
	
	\begin{subfigure}[t]{.26\textwidth}
		\includegraphics{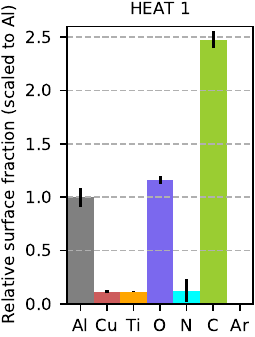}
	\end{subfigure}%
	\hskip 1ex
	\begin{subfigure}[t]{.18\textwidth}
		\includegraphics{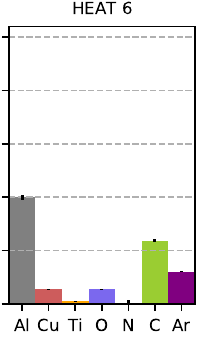}
	\end{subfigure}%
	
	\vspace{.3 cm}

	\hskip 1ex
	\begin{subfigure}[t]{.26\textwidth}
		\includegraphics{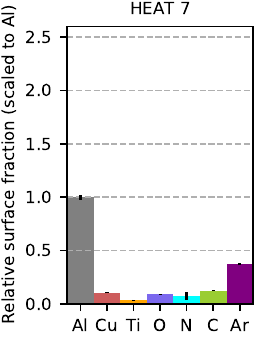}
	\end{subfigure}%
	\hskip 1ex
	\begin{subfigure}[t]{.18\textwidth}
		\includegraphics{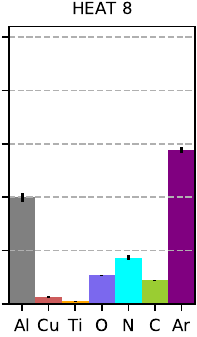}
	\end{subfigure}%
	
	\vspace{.3 cm}
	
	\hskip 1ex
	\begin{subfigure}[t]{.26\textwidth}
		\includegraphics{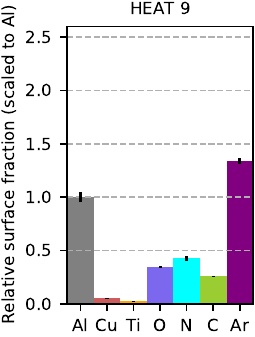}
	\end{subfigure}%
	\hskip 1ex
	\begin{subfigure}[t]{.18\textwidth}
		\includegraphics{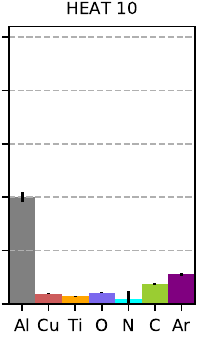}
	\end{subfigure}%
	
	\vspace{0cm} 
	\caption{Relative fractional composition of the trap surface during six heat treatments, scaled such that the aluminum fraction is constant. The trap was milled with argon ions between heat treatments, causing the surface composition to change. These surface composition measurements were completed immediately before the heat treatments indicated, with the exception of the \textsc{heat 9} measurement, which took place immediately after.  The Auger spectra from which this data was extracted can be found in App.~\ref{ap:auger_raw}.} 
	\label{fig:heat_rep_surface_frac}
	\end{figure}

Surface compositions corresponding to six of the ten heat treatments are shown in Fig.~\ref{fig:heat_rep_surface_frac}. The corresponding Auger measurements were performed immediately before treatments \textsc{heat 1, heat 6, heat 7, heat 8} and \textsc{heat 10}, and immediately after \textsc{heat 9}. The Auger spectra from which these surface compositions were extracted can be found in App. \ref{ap:auger_raw}. The surface composition was not measured immediately before or after treatments \textsc{heat 2, 3, 4} or 5. 

\begin{figure}[ht]
	\centering
	\includegraphics{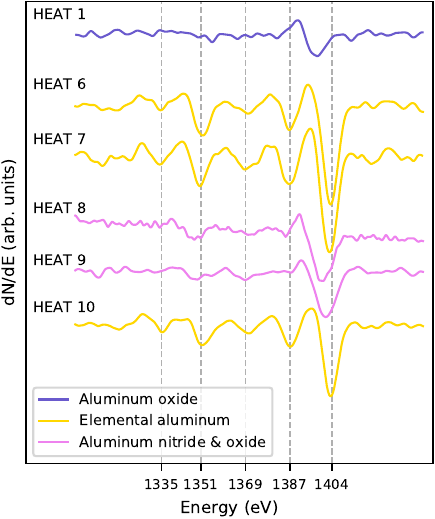}
	\caption{Measured aluminum Auger spectra. The aluminum line-shape changed as the surface was milled with argon ions. These Auger spectra were taken immediately before the heat treatments indicated, with the exception of the \textsc{heat 9} measurement, which was taken immediately after. }
	\label{fig:aluminum_lineshape}
\end{figure}

In addition to containing information about the atomic composition of the trap, Auger spectra also allow us to determine the chemical state of the aluminum at the trap surface. The aluminum state can be deduced from the lineshape of the aluminum Auger peaks, combined with the ratio of aluminum to other elements. Aluminum Auger spectra taken before or after six of the ten heat treatments are shown in Fig.~\ref{fig:aluminum_lineshape}.

The ratio of oxygen to aluminum at the trap surface was greater than 1:1 during \textsc{heat 1}, as shown in Fig.~\ref{fig:heat_rep_surface_frac}. This, in combination with a shift of the 1404~eV Auger peak by -8~eV, as shown in Fig.~\ref{fig:aluminum_lineshape}, is indicative of oxidized aluminum \cite{Timmermans2002-Al-chem-auger}. This is consistent with the trap's history, as it had been exposed to atmosphere but not yet milled at the start of \textsc{heat 1}. The presence of carbon suggests that hydrocarbon contamination also built up on the surface during this exposure.

The aluminum spectra measured before \textsc{heat} 6, 7, and 10, and plotted in Fig.~\ref{fig:aluminum_lineshape}, each have five distinct peaks and qualitatively match spectra of elemental aluminum \cite{McGuire1979-auger} ~\footnote{We note that the largest feature in these measured spectra can be found at 1404~eV, while the largest feature in the elemental aluminum reference spectrum in \cite{McGuire1979-auger} is located at 1389~eV. We attribute this mismatch to electron-analyzer calibration differences.}. The corresponding surface compositions plotted in Fig.~\ref{fig:heat_rep_surface_frac} show that the surface contamination was relatively low during these heat treatments, which is consistent with the aluminum being primarily elemental.

The aluminum spectra measured before \textsc{heat} 8 and after \textsc{heat} 9, are plotted in Fig.~\ref{fig:aluminum_lineshape}. The small negative energy shift of the primary peak 1404~eV, and the reduction of the satellite peaks at 1135, 1369 and 1387~eV, are indicative of aluminum nitride \cite{Timmermans2002-Al-chem-auger}. The corresponding surface fractions, shown in Fig.~\ref{fig:heat_rep_surface_frac}, show significant levels of both oxygen and nitrogen.  We conclude that during these heat treatments, the aluminum surface was partially oxidized, and partially took the form of aluminum nitride. It is unclear whether the oxygen and nitrogen were embedded in the bulk of the metal and then revealed by milling, or whether these elements were introduced to the trap surface during the ion milling process.

\begin{figure}[ht]
	\centering
	\begin{subfigure}[t]{.21\textwidth}
		\includegraphics{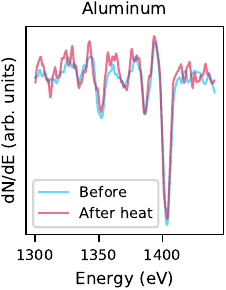}
	\end{subfigure}%
	\hskip 2ex
	\begin{subfigure}[t]{.19\textwidth}
		\includegraphics{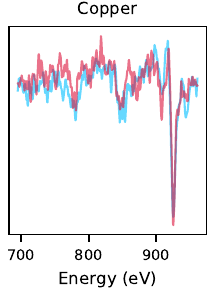}
	\end{subfigure} \\
    \vspace{.5 cm}
	\begin{subfigure}[t]{.21\textwidth}
		\includegraphics{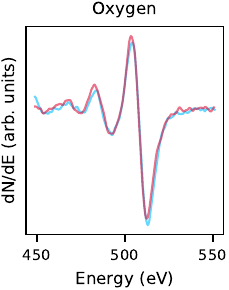}
	\end{subfigure}%
	\hskip 2ex
	\begin{subfigure}[t]{.19\textwidth}
		\includegraphics{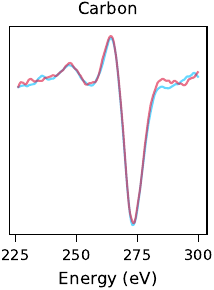}
	\end{subfigure}%
	
	\caption{Auger spectra measured before and after thermal transformations: aluminum in \textsc{heat 6},  copper in \textsc{heat 10}, oxygen in \textsc{heat 10}, carbon in \textsc{heat 7}. No thermally-driven line-shape changes are visible. Lines measured before and after heat treatments have been scaled to the magnitude of each element's largest peak to enable a direct comparison of the overall shape of the spectrum of that element.} \label{fig:four_aug_lineshapes}
\end{figure}

We compare the aluminum Auger spectra measured before and after \textsc{heat 6}, when a significant thermal transformation took place, in Fig. \ref{fig:four_aug_lineshapes}. We do not observe any significant changes in the aluminum lineshape, indicating that the elemental aluminum surface did not oxidize during this thermal transformation. Also included in Fig.~\ref{fig:four_aug_lineshapes} are plots of copper, oxygen and carbon Auger line-shapes measured before and after thermal transformations. We do not observe qualitative changes in any of these lineshapes.

\begin{figure}[ht]
	\centering
	\begin{subfigure}[t]{.48\textwidth}
		\includegraphics{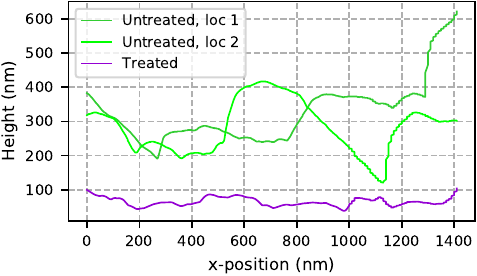}
	\end{subfigure} \\
	\vspace{.2 cm}
	\begin{subfigure}[t]{.48\textwidth}
		\includegraphics{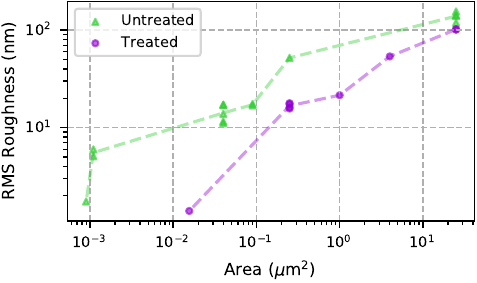}
	\end{subfigure}%
	\caption{Cross sections and RMS roughness values corresponding to two metal films. These measurements were performed using atomic force microscopy (AFM). The \textit{Treated} sample refers to ion trap discussed throughout this manuscript. The \textit{Untreated} film was fabricated using the same methods as the ion trap, and never milled or heated above room temperature. The RMS roughness of the untreated trap is generally higher than that of the treated trap.}
	\label{fig:afm_data}
\end{figure}

\paragraph*{\textit{Ex situ} AFM imaging:}
The ion trap was removed from vacuum after \textsc{heat 10} and imaged using atomic force microscopy (AFM). The results of these measurements are plotted in Fig~\ref{fig:afm_data}. As this trap had undergone extensive heat treatments and argon ion milling previous to these measurements, it is referred to in these plots with the label \textit{Treated}. The cross section and RMS roughness of this trap are compared to those of an untreated thin film which was fabricated in an identical manner. This film was stored in atmosphere for several years before the AFM measurements were performed, but it was not milled or heated above room temperature after fabrication. The RMS roughness of the treated film is lower than that of the untreated film on scales smaller than 0.25~$\mu$m$^2$. This is consistent with the comparatively large features that are present in the cross sections of the untreated film, and absent from the cross section of the treated trap, as shown in Fig.~\ref{fig:afm_data}.

\section{Analysis} \label{sec:analysis}

The parameters of the ten heat treatments discussed in this work, as summarized in Tab.~\ref{ch:heat_tab:heat}, are different for each treatment. However, during each heat treatment, the trap was heated to temperatures above 555~K for two or more days, or above 575~K for one or more days. Considering the dynamics of the thermal transformation during the experiments plotted in Fig.~\ref{fig:m7_v_time_recrystal} and Fig.~\ref{fig:m12_v_time_recrystal}, we expect that each of the ten heat treatments would have generated a thermal transformation had the surface been primed for one. The observed variations in the thermal transformation magnitudes, as plotted in Fig.~\ref{fig:recrystal_all_tempscalings}, cannot be explained by variations in heat treatment parameters.

\begin{figure*}[ht!]
	\centering
		\includegraphics{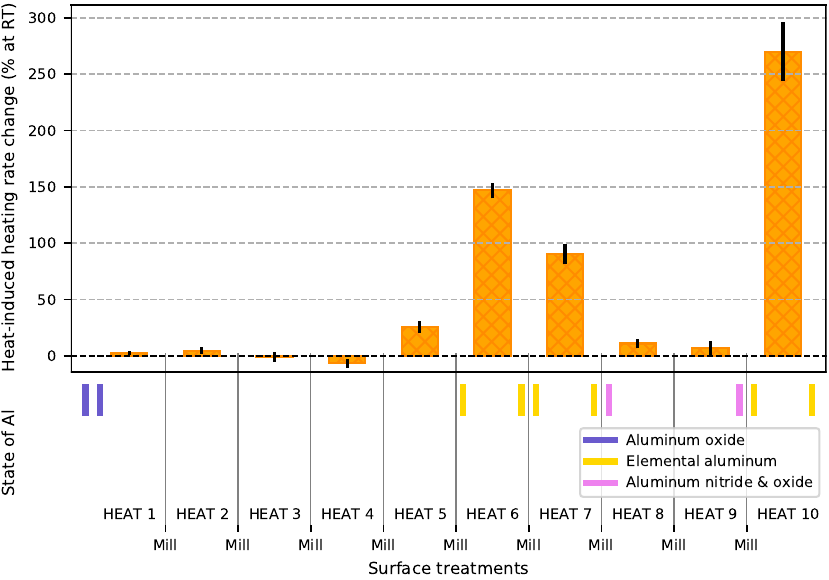}
	        \caption{Data from heating rates measured at room temperature before and after heat treatments, shown in context with the chemical state of the aluminum on the trap surface as deduced from measured Auger spectra. Orange bars indicate the measured fractional change of the room-temperature heating rate in response to various heat treatments. Yellow, purple and pink bars indicate the chemical state of aluminum as determined from the Auger spectra. } \label{fig:percent_change_surffrac_hr}
\end{figure*}

We can investigate why thermal transformations took place during some heat treatments and not others by studying the information contained in the measured Auger spectra. From the surface compositions presented in Fig.~\ref{fig:heat_rep_surface_frac}, and the aluminum lineshapes presented in Fig.~\ref{fig:aluminum_lineshape}, it is evident that the composition of the trap surface was constantly changing throughout the surface treatment experiments. We  now compare the evolving chemical state of the aluminum at the surface to the presence and absence of thermal transformations. These features are shown together in Fig.~\ref{fig:percent_change_surffrac_hr}.

Thermal transformations did not take place during early heat treatments when the aluminum on the surface of the trap was oxidized, as shown in Fig.~\ref{fig:percent_change_surffrac_hr}. As the trap was stored in atmosphere both before and after baking for a cumulative exposure time of 30 weeks, the oxide on the aluminum-copper surface would have been between 2 and 10~nm thick at the start of our experiments. The transformations first occurred during \textsc{heat 5}, by which point milling had removed a total of 6.3~nm from the surface, likely breaking through some of the oxide layer to reveal elemental aluminum. During treatment \textsc{heat 6}, the surface was primarily composed of elemental aluminum, and the subsequent thermal transformation was significant. Later, when elemental aluminum was replaced by aluminum nitride and aluminum oxide before \textsc{heat 8}, the thermal transformations decreased in magnitude and then ceased entirely. Finally, aggressive angled milling before \textsc{heat 10} removed the nitride-oxide layer and re-exposed elemental aluminum, at which point the thermal transformation returned. 

Since the occurrence of thermal transformations is correlated with the chemical state of the aluminum at the trap surface, we consider physical processes that are consistent with this pattern. Processes including contaminant deposition, surface oxidation, compound growth, and recrystallization of the metal all could take place in a system such as ours. In addition, the specifics of these processes would all be affected by the oxidation or nitrogenation state of the metal surface. We will now discuss each of these mechanisms in turn, and consider whether any of them can be driving the observed thermal noise transformations.

\paragraph*{Contamination level:}
Non-metal contaminants are thought to be a major producer of surface electric-field noise, so we first consider whether thermal transformations could be driven by a rise in the contamination level of the trap surface. Elements in the vacuum chamber close to the ion trap heater outgas during heat treatments, and some of these gases may adsorb more readily to elemental aluminum than to aluminum oxide or nitride. Heat can also cause certain elements, in particular carbon, to diffuse from the bulk to the surface of a metal. These elements may diffuse more readily through elemental aluminum than through aluminum oxide or nitride.

With the exception of hydrogen contamination, a rise in the contaminant level at the trap surface during a thermal transformation should be observable with Auger spectroscopy. When we compare spectra measured before and after a thermal transformation, as shown in Fig.~\ref{fig:auger_spectra_recrystal}, we do not observe signals from any new elements arising after the transformation. In addition, contaminant levels compared before and after three transformations, as shown in Fig.~\ref{fig:surf_frac_changes_line_recrystal}, do not exhibit any significant trends. From these measurement results we conclude that the thermal noise transformations are not driven by changes in the level of (non-hydrogen) contamination near the trap surface.

\paragraph*{Chemistry at the trap surface:}
Although there are no systematic shifts in the elemental composition of the trap surface during thermal transformations, the heat treatments could be driving changes in the chemical structure of the existing elements. Whether the metal surface participates in these reactions as a catalyst or as a reactant, the oxidation or nitrogenation state of the metal could have a major impact on the active chemistry. 

Some chemical information can be deduced from the lineshapes of Auger spectra. Thus, a change in one of these lineshapes would indicate that a chemical reaction had taken place. In Fig.~\ref{fig:four_aug_lineshapes}, we plot aluminum, copper, oxygen and carbon Auger spectra before and after thermal transformations, and find no change in the lineshapes. In particular, we note that the elemental aluminum does not become oxidized or nitrogenated during the heat treatment. 

The observed stability of the Auger lineshapes does not rule out chemical changes in general, however, as some chemistry is not visible with Auger spectroscopy. For example, when the incident electron beam from an Auger spectrometer hits the surface of a material, the electrons break up some chemical compounds before their energy signatures can be measured. Hydrocarbon structures are particularly sensitive to energetic electrons \cite{Pantano1981}. As a result, even if hydrocarbon growth was the driver of a thermal transformation, we might not see evidence of this in any Auger line-shapes. However, if hydrocarbon compound growth drove a rise in ion heating rates, then bombarding these compounds with energetic electrons should cause the heating rates to go back down. We performed electron bombardment after a thermal transformation, see Fig.~\ref{fig:electron4}. The transformation is not reversed by electron bombardment, and thus we conclude that hydrocarbon growth cannot explain the thermally-driven increase of ion heating rates. Instead, electron bombardment slightly increased ion heating rates at high temperatures, which we speculate was caused by the electron-beam-induced deposition of carbon.

\paragraph*{Structural changes of the metal film:}
We now shift our focus to the arrangement of atoms in the film. If atoms in elemental aluminum are mobile at our heat treatment temperatures of 550 to 600~K, and atoms in aluminum oxide and nitride are rigid at these temperatures, then a restructuring of the metal can explain why thermal noise transformations only take place when elemental aluminum is present at the surface. 

The recrystallization temperatures of materials similar to ours can provide insight into the temperatures at which atoms in our trap surface become mobile. The \emph{initial recrystallization temperature} of a material is the temperature at which small crystal grains start to combine to form larger grains on a timescale of hours. The initial recrystallization temperature of bulk aluminum is 420~K \cite{askeland_science_2011}. In a set of experiments performed with aluminum foils, Drits \textit{et al.} \cite{Drits1971} found that the addition of impurities raises the recrystallization temperature of elemental aluminum, with the effect saturating as the concentration of impurities is raised. In particular, they found that the addition of titanium raises the recrystallization temperature of aluminum by 110~K and saturates at an impurity concentration of 0.5\%. They separately found that copper raises the recrystallization temperature by 30~K. 

The noise transformation observed during treatment \textsc{heat 7}, see Fig.~\ref{fig:m7_v_time_recrystal}, saturated with a half life of 1.6$\pm$0.8 days at 575~K. During this heat treatment, the trap surface was primarily composed of elemental aluminum, with small amounts of copper, titanium, oxygen, nitrogen, carbon and argon present as well. With titanium alone, we can expect the bulk recrystallization temperature to rise to 530~K, and additional impurities may further raise the recrystallization temperature of the film by a small amount. The expected recrystallization temperature range is consistent with the observed temperatures and timescales of thermal noise transformations. In contrast, the recrystallization temperatures of aluminum oxide and nitride are significantly higher than the temperatures reached in our experiments. An amorphous thin film of aluminum oxide crystallizes at 720 to 770~K \cite{Qin2014} while an amorphous film of aluminum nitride crystallizes at 1300~K \cite{broas_structural_2016}. 

Atomic rearrangement may occur at different temperatures at the surface compared to the bulk, as atomic mobility can be higher at surfaces \cite{Rhead1989}. The dynamics of our film are further complicated by surface roughness, partial oxidation, and unique properties of metal films deposited on glass substrates. Even considering this uncertainty, atoms in our elemental aluminum film are expected to be significantly more mobile than atoms in aluminum oxide and aluminum nitride, at the temperatures explored in this study. Thus, atomic restructuring could explain why thermal transformations only took place when the trap surface was composed of elemental aluminum, rather than aluminum oxide or nitride.

The structure of annealed metal depends not only on the annealing temperature and time, but also on the rate at which the temperature of the metal is cycled. In addition, some metals exhibit secondary recrystallization, where a partial recrystallization plateaus at one temperature, and then continues recrystallizing after a higher temperature threshold is reached~\cite{Thompson1990-film-grains}. These behaviors could explain why, during heat \textsc{heat 7} and 10, the thermal transformation appeared to saturate, but then continued to rise after subsequent temperature cycling and high-temperature heat treatment, as shown in Fig.~\ref{fig:m12_v_time_recrystal}. 

\paragraph*{Thermally activated fluctuators:}

The first ion heating-rate measurements performed in this trap were consistent with noise from an ensemble of thermally activated fluctuators (TAFs), as detailed in a previous publication \cite{Noel2019-TAF}. The TAF model does not specify a physical noise mechanism. Rather, it is a mathematical framework which can provide insight into the dynamics of physical noise sources. In the context of this trap, TAF noise is driven by thermally activated jumps between semi-stable states with different charge configurations. The noise spectrum of an individual TAF is determined by the activation energy of this jump. 

In a macroscopic physical system, such as the surface of an ion trap, the measured TAF noise will include contributions from many different fluctuators. TAF noise at a fixed temperature and frequency is dominated by a subset of TAFs with a specific range of activation energies. Thus by measuring ion heating rates at different temperatures, one can map out the relative magnitude of noise from different TAFs. 

The temperature dependence of noise generated by a distribution of TAFs will never include features sharper than the width of an individual fluctuator, as illustrated in \cite{Noel2019-TAF}. Beyond this, the TAF model does not predict a functional form for how noise from a generic distribution of fluctuators should scale with temperature. Instead, this model establishes that there is a specific relationship between the temperature scaling and frequency scaling of TAF noise. Thermally activated jumps generate zero-frequency Lorentzian noise with a bandwidth corresponding to the jump rate. Thus, raising the temperature of the system also raises the frequency of the noise.

As we investigate thermal transformations, we are interested in identifying the energy scale of relevant microscopic noise processes. Such insight might be gleaned by considering how the TAF distribution changes during these transformations. Before performing this analysis, we note that although the initial measurements in this trap were consistent with TAF noise, we cannot assume that this consistency persisted after extensive surface treatment. To evaluate whether the TAF model can describe the noise at any given stage of the experiment, we need to verify its predictions. We can do so by measuring the temperature scaling of the electric field noise at multiple frequencies.

We measured ion heating rates as a function of temperature at two frequencies both before and after only three of the observed thermal transformations: those that took place during \textsc{heat 6, 7} and the second part of \textsc{heat 10}. To evaluate the applicability of the TAF model to these data sets, we analyze the slope of each heating-rate temperature dependence in the context of the TAF model to predict the temperature dependence of frequency scaling exponent $\alpha$. App.~\ref{app:taf} describes this analytical process in more detail.

\begin{figure}[ht!]
	\centering
	\begin{subfigure}[t]{.27\textwidth}
		\includegraphics{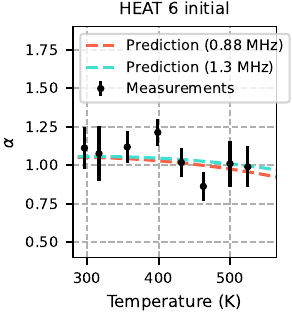}
	\end{subfigure}
	\begin{subfigure}[t]{.20\textwidth}
		\includegraphics{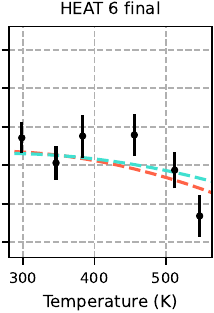}
	\end{subfigure} \\
	\vspace{.2cm}
	\begin{subfigure}[t]{.27\textwidth}
		\includegraphics{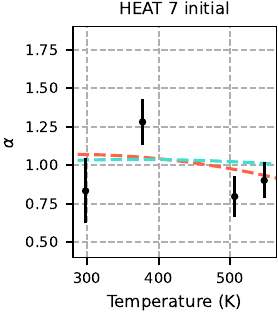}
	\end{subfigure}
	\begin{subfigure}[t]{.20\textwidth}
		\includegraphics{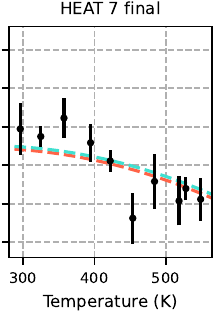}
	\end{subfigure} \\
	\vspace{.2cm}
	\begin{subfigure}[t]{.27\textwidth}
		\includegraphics{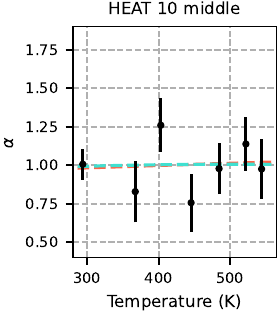}
	\end{subfigure}
	\begin{subfigure}[t]{.20\textwidth}
		\includegraphics{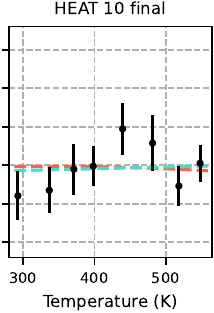}
	\end{subfigure}
	\caption{Direct measurements of the temperature scaling exponent $\alpha$ compared to predictions from the TAF model. These predictions are extracted from the slopes of the heating-rate temperature dependencies measured at 2$\pi \times 0.88$ and 2$\pi \times 1.3$~MHz, as described in App.~\ref{app:taf}.}
	\label{app:fig:all_alpha}
\end{figure}

Measurements of $\alpha$ at each temperature are extracted from noise measurements at 2$\pi \times 0.88$ and 2$\pi \times 1.3$~MHz, using Eq.~\ref{eqn:freq_scale}. The extracted $\alpha$ values from these data sets are compared to the $\alpha$ predictions, as plotted in Fig.~\ref{app:fig:all_alpha}. 

To quantify the accuracy of the TAF prediction, we perform a $\chi^2$ analysis comparing the measured values of $\alpha$ to the predictions from the TAF model extracted from the Gaussian fit of the 2$\pi \times 1.3$~MHz heating-rate temperature dependence, as described in App.~\ref{app:taf}. The p-value of this data-set is determined from the reduced $\chi^2$ value $\chi^2_r$, using the number of $\alpha$ measurements $n$ as the number of degrees of freedom. This analysis is repeated with a comparison of the measured $\alpha$ values to the $\alpha = 1$ null hypothesis. We note that both of these are zero-parameter fits; neither the prediction from the TAF model nor the $\alpha = 1$ null hypothesis depend on the measured values of $\alpha$. 
 
\begin{center} 
\begin{table}[h!]
  \begin{center}
	  \small
    \caption{Quantitative analysis of fit to thermally activated fluctuator model and to the $\alpha$ = 1 null hypothesis.}
    \label{tab:pvalues_taf}
    \begin{tabular}{| l | r | r |r | r | r | r | r }
      \hline
      \textbf{Treatment}    & \textbf{$\#$ of}   & \textbf{$\boldsymbol\chi^2_r$}  & \textbf{$\boldsymbol\chi^2_r$}  & \textbf{p-value}  & \textbf{p-value}    \\    
                            & \textbf{data}      & \textbf{TAF}                  & \textbf{$\boldsymbol\alpha$=1}& \textbf{TAF}      & \textbf{$\boldsymbol\alpha$=1}    \\ 
                            & \textbf{points}    &                               &                               &                   &                     \\ 

      \hline
	  \textsc{heat} 6 initial$\ $                     &   8    &   0.86  &  1.23   &   0.55  & 0.28  \\  
	  \textsc{heat} 6 final$\ \ \ $                   &   6    &   1.25  &  2.17   &   0.28  & 0.04  \\  
	  \textsc{heat} 7 initial$\ $                     &   4    &   1.85  &  1.80   &   0.12  & 0.13  \\  
	  \textsc{heat} 7 final $\ \ \ $                  &   10   &   1.05  &  3.05   &   0.40  & 0.001 \\ 
      \textsc{heat} 10 middle                         &   7    &   0.76  &  0.76   &   0.62  & 0.62  \\ 
	  \textsc{heat} 10 final$\ \ \ $                  &   8    &   0.73  &  0.80   &   0.67  & 0.60  \\  
      \hline   
    \end{tabular}
	\normalsize
  \end{center}

\end{table}
\end{center}

Our analysis of data taken before and after three thermal transformations shows that these six data sets fit to the TAF model with p-values greater than 0.1, as shown in Tab.~\ref{tab:pvalues_taf} and Fig.~\ref{app:fig:all_alpha}. Furthermore, in these six cases the p-value of the TAF model is similar or greater than the p-value of the $\alpha = 1$ null hypothesis, as shown in Tab.~\ref{tab:pvalues_taf}.

\begin{figure}[ht!]
	\centering

	\begin{subfigure}[t]{.19\textwidth}
		\includegraphics{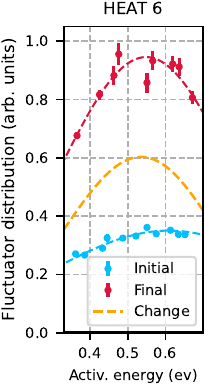}
	\end{subfigure}
	\begin{subfigure}[t]{.145\textwidth}
		\includegraphics{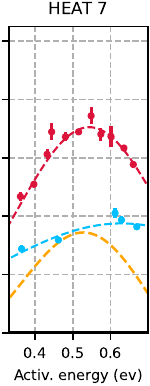}
	\end{subfigure}%
	\begin{subfigure}[t]{.14\textwidth}
		\includegraphics{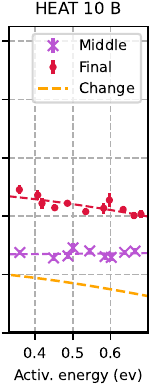}
	\end{subfigure}%

	\caption{TAF distributions extracted from ion heating rates measured at a range of different temperatures before and after thermal transformations. We measure noise at frequencies near 1~MHz, and at trap temperatures between 290 and 600~K, so we can map out distributions of TAFs with activation energies from 0.35 to 0.7~eV. The relative magnitude of the fluctuator distribution at each activation energy depends on the product of the density and noise magnitude of the TAFs with that energy.
	} 
	\label{fig:recrystal_taf_heat}
\end{figure}

Having established that the TAF model is relevant to thermal transformations \textsc{heat 6, 7} and 10~\textsc{b}, we proceed with our discussion of these transformations in the context of the TAF model. Using the procedure described in Ref.~\cite{Noel2019-TAF}, and ion heating rates measured as a function of temperature, we map out the TAF distributions corresponding to the noise before and after each transformation. These distributions are plotted in Fig.~\ref{fig:recrystal_taf_heat}. In all three cases, the thermal transformation caused the magnitude of the TAF distributions to increase at all measured activation energies. An increase in the fluctuator distribution corresponds to a rise in the density or strength of fluctuators at a given activation energy. In \textsc{heat 6} and \textsc{heat 7}, TAFs with activation energies around 0.55~eV had the greatest increase. In \textsc{heat 10 b}, TAFs at lower activation energies had the greatest increase. 

Although we focus here on on the TAF model as it pertains to changes that take place during thermal transformations, we also collected data at additional stages throughout the full sequence of experiments performed on this trap, in order to evaluate the efficacy of the TAF model in many different contexts. Some additional measurements performed in this trap gave p-values below 0.02 for both the TAF model and the $\alpha = 1$ null hypothesis. This data has been fully presented and discussed elsewhere \cite{Berlin-Udi2020TreatmentsSurface}. To conclude our discussion of the TAF model, we also acknowledge that there may be other models that can describe some of our observations.

\section{Discussion \& Conclusion} \label{sec:discussion}

In a series of experiments performed on a single ion trap, we observed that under certain conditions, ion heating rates can rise as a result of an \textit{in situ} heat treatment. In an effort to illuminate the underlying physics of this thermal transformation, we studied and characterized this process using ion milling, electron bombardment, heat treatments, Auger spectroscopy, and ion heating rate measurements. To characterize the timescale and temperature dependence of the transformation, we held the trap at elevated temperatures while repeatedly measuring ion heating rates. In one experiment we found the transformation to saturate on a timescale of 1.6 days at 575~K. Secondary transformations were observed after subsequent temperature cycling, and when the trap was heated above 605~K. 

Argon ion milling altered the composition of the trap surface, and we tracked these changes with Auger spectroscopy. As the trap's surface was milled, the ion heating rate steadily dropped. This drop may be partially explained by a reduction of the contamination level near the surface of the trap as material was removed. We also found a correlation between the behavior of thermal transformations and the chemical state of the aluminum on the trap surface: when the aluminum was in the elemental form, the noise transformed in response to a heat treatment. When the aluminum was oxidized or nitrogenized, no transformation took place. Auger spectra taken before and after thermal transformations showed that thermal transformations were not accompanied by a detectable rise in contamination levels, and the surface also did not oxidize during heat treatments. Electron bombardment treatments allow us to rule out complex hydrocarbon structure growth as the driver of thermal transformations: complex hydrocarbons are highly susceptible to electron damage, and electron bombardment did not counteract the effects of thermal transformations. We cannot rule out changes in the hydrogen content of the surface, as Auger spectroscopy is not sensitive to hydrogen.

The observed temperatures and saturation behaviors of the noise transformations are consistent with the rearrangement of atoms near the surface of the metal. Transformations have only taken place when the aluminum at the trap surface was in the elemental state, as opposed to a state of oxidation or nitrogenation. Atoms in aluminum nitride and aluminum oxide are far less mobile than those in elemental aluminum at temperatures between 500~K and 600~K, as evidenced by their respective recrystallization temperatures. These differences in atomic mobility, combined with the observation that the transformations only take place when the aluminum at the surface is elemental, are consistent with atomic restructuring as a possible driver of thermal noise transformations. 

Milling and heating act in opposition in our system, with milling both reversing the effects of thermal noise transformations, and also priming the surface for thermal transformations to take place. There are many different ways in which heating and ion milling could be acting in opposition, including by altering the crystallinity of the metal, the roughness of the surface, or the arrangement of different elements within the film. Any of these processes, which are explored in more depth in App.~\ref{ap:restructuring}, could be affecting electric-field noise in this ion trap. 

In a previous publication, we demonstrated that the noise in this trap was consistent with a distribution of thermally activated fluctuators (TAFs) \cite{Noel2019-TAF}. Here, we showed how the TAF distribution rose in response to thermal transformations. With our experimental setup, we can map out noise from TAFs with activation energies between 0.35 and 0.7~eV. Resistance noise in polycrystalline metal films has been found to be consistent with TAFs that peak in a similar range: between 0.5 and 1~eV \cite{Eberhard1978ExcessMetals,Dutta1981-review}. This resistance noise has been linked to TAFs manifesting as defects fluctuating at grain boundaries \cite{Dutta1981-review, Weissman19881fMatter,Zhigalskii1997Films}. The similarities between our electric-field noise measurement results and these resistance noise measurement results indicate that the noise in our trap may be linked to fluctuating defects in the film. This lends credence to the hypothesis that thermal noise transformations may be caused by a restructuring of the metal film and its impurities.

Our results suggest that it will be important to explore the impact of film structure, surface morphology, and defect dynamics on surface electric-field noise through experimentation with ion traps. Such experiments may include measuring ion heating rates in a thermally transformed trap below room temperature, performing \textit{in situ} annealing experiments on ion traps with different recrystallization temperatures, and building ion traps with nano-structured surfaces.

\begin{acknowledgments}
The authors acknowledge Adam Udi for assistance with software development. Part of the trap fabrication was performed in the UC Berkeley Marvell Nanofabrication Laboratory. M.B.-U. acknowledges an NSF Graduate Research Fellowship. A.M. A, B.S., H.H. were supported in part by the  U.S. Department of Energy via grant DE-SC0019376 \&
LBNL QSA Center. Part of this  work  was  performed  under  the  auspices of  the  U.S.  Department  of  Energy  by  Lawrence  Livermore National Laboratory under Contract DE-AC52-07NA27344.
\end{acknowledgments}

\bibliography{references_manual.bib}

\newpage

\appendix

\section{Auger spectroscopy} \label{ap:auger_raw}

\begin{figure}[ht!]
	\begin{subfigure}[t]{.185\textwidth}
		\includegraphics{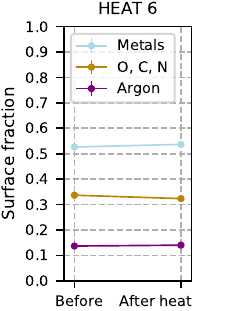}
	\end{subfigure}%
	\begin{subfigure}[t]{.14\textwidth}
		\includegraphics{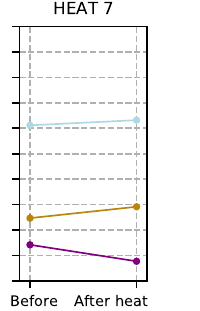}
	\end{subfigure}%
	\begin{subfigure}[t]{.16\textwidth}
		\includegraphics{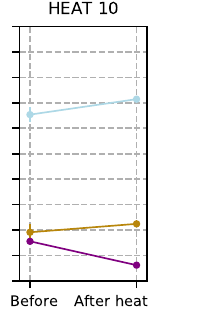}
	\end{subfigure} \\
	\vspace{.5 cm}
	\begin{subfigure}[t]{.185\textwidth}
		\includegraphics{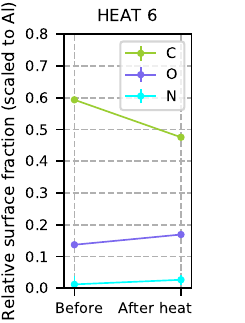}
	\end{subfigure}%
	\begin{subfigure}[t]{.14\textwidth}
		\includegraphics{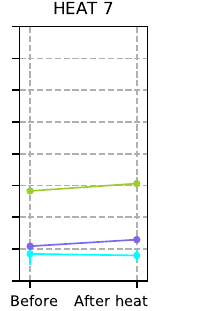}
	\end{subfigure}%
	\begin{subfigure}[t]{.16\textwidth}
		\includegraphics{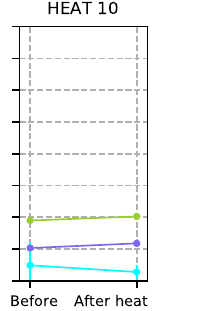}
	\end{subfigure}%
	\caption{Surface fractions measured before and after thermal transformations. Metals includes aluminum, copper and titanium. In the lower three plots, carbon, oxygen and nitrogen are scaled relative to aluminum. We observe no statistically significant trends.} \label{fig:surf_frac_changes_line_recrystal}
\end{figure}

In comparing Auger spectra measured before and after argon ion milling, it is evident that the composition of the ion trap surface changed significantly as a result of milling. In contrast, the relative peak magnitudes were much less affected by heat treatments. The relative surface fractions of identified elements are compared before and after three heat treatments in Fig.~\ref{fig:surf_frac_changes_line_recrystal}. With the possible exception of a drop in Argon level, there are no consistent, statistically significant detectable trends in the surface composition. Small variations between measurements are expected as each scan was performed on a slightly different location on the substrate. 

\begin{figure*}[ht!]
	\centering
		\includegraphics[scale=1]{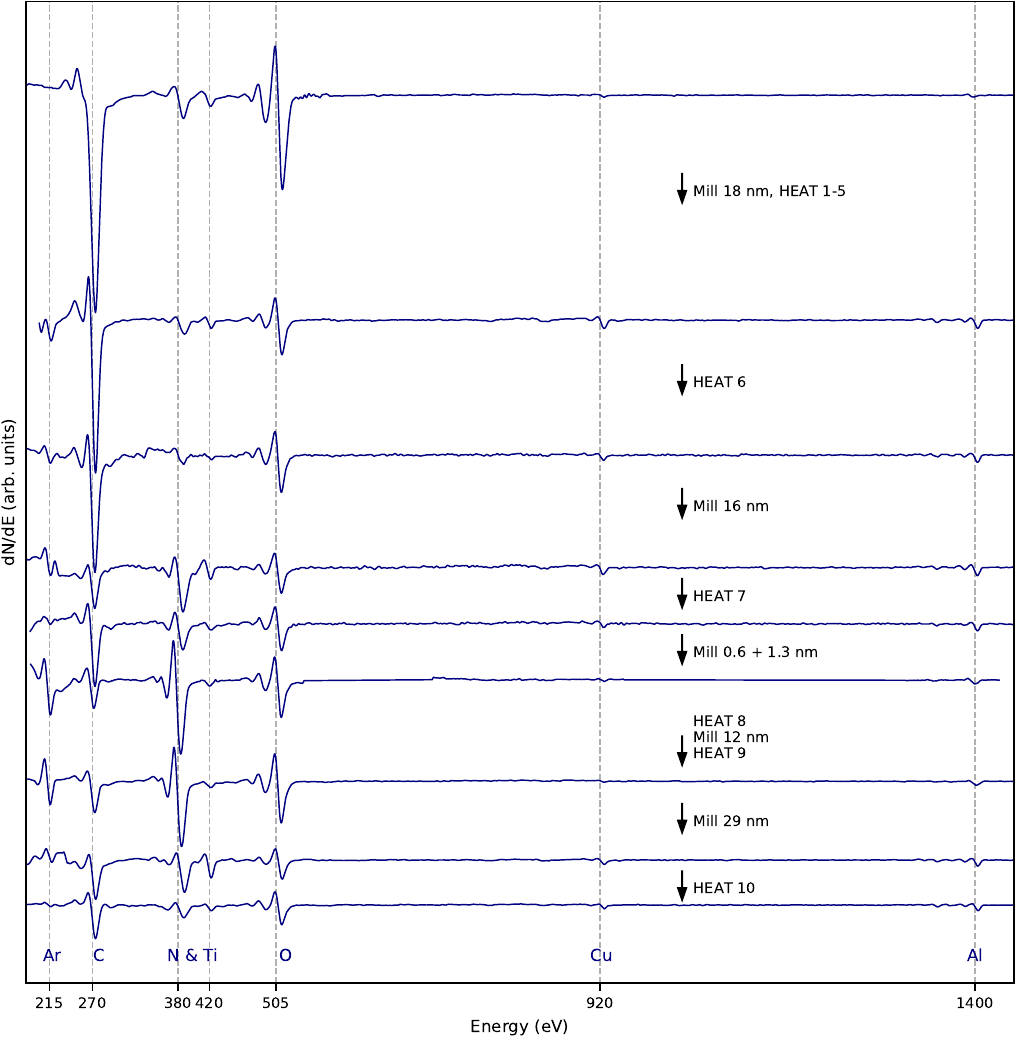}
	        \caption{Data from Auger spectroscopy measurements performed \textit{in situ} on the ion trap. Before plotting and analysis was performed, large-scale magnitude drifts were subtracted out. Features identified as characteristic peaks of specific elements are labeled. Notes on heating and milling describe the surface treatments that were performed between Auger spectroscopy measurements. } \label{fig:auger_raw}
\end{figure*}

We reported, in Figs.~\ref{fig:heat_rep_surface_frac} and \ref{fig:surf_frac_changes_line_recrystal}, the elemental composition of the ion trap surface at various times throughout a series of surface treatment experiments. The Auger spectra from which these compositions were extracted are presented in Fig.~\ref{fig:auger_raw}. All measurements were performed \textit{in situ}, as in vacuum was not broken between surface treatments and Auger spectroscopy measurements.

\section{Surface treatments} \label{ap:history}

\begin{figure*}[ht!]
	\centering
	\begin{subfigure}[t]{.28\textwidth}
		\includegraphics{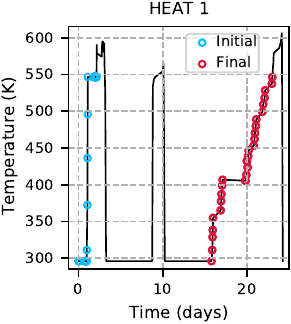}
	\end{subfigure}
	\begin{subfigure}[t]{.22\textwidth}
		\includegraphics{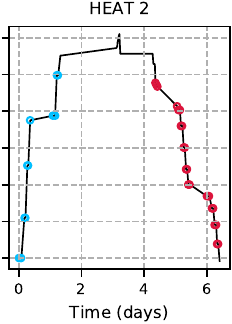}
	\end{subfigure}
	\begin{subfigure}[t]{.22\textwidth}
		\includegraphics{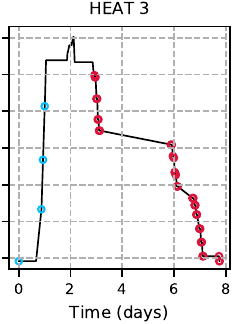}
	\end{subfigure}
	\begin{subfigure}[t]{.22\textwidth}
		\includegraphics{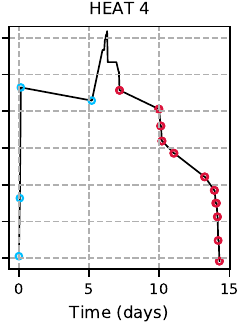}
	\end{subfigure}

	\vspace{.3cm}
	
	\begin{subfigure}[t]{.28\textwidth}
		\includegraphics{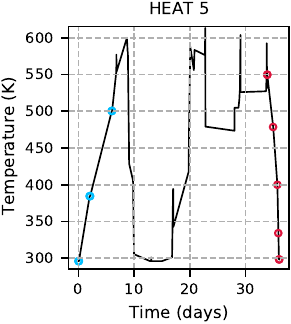}
	\end{subfigure}
	\begin{subfigure}[t]{.22\textwidth}
		\includegraphics{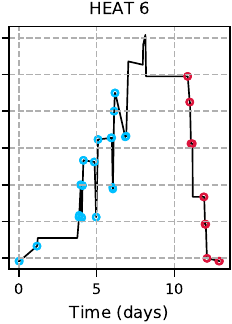}
	\end{subfigure}
	\vspace{.3cm}
	\begin{subfigure}[t]{.22\textwidth}
		\includegraphics{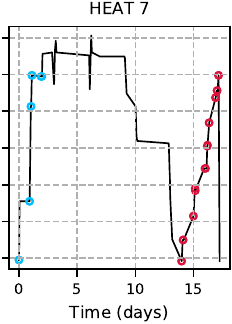}
	\end{subfigure}
	\begin{subfigure}[t]{.22\textwidth}
		\includegraphics{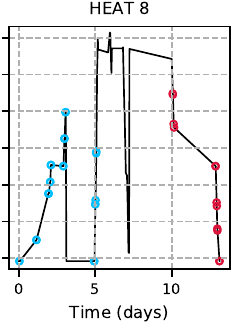}
	\end{subfigure}

	\vspace{.3cm}

	\begin{subfigure}[t]{.28\textwidth}
		\includegraphics{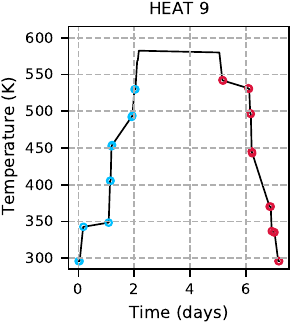}
	\end{subfigure}
	\begin{subfigure}[t]{.22\textwidth}
		\includegraphics{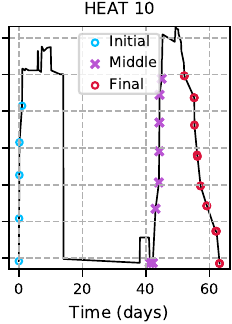}
	\end{subfigure}
	\caption{The timing and temperatures of ten heat treatments performed on the ion trap. In Fig. \ref{fig:recrystal_all_tempscalings} ion heating rates were reported as a function of temperature, and labelled as \textit{initial}, \textit{middle}, or \textit{final}. The points marked here, and labeled in the same manner, indicate in more detail the context surrounding the ion heating rate measurements reported in Fig. \ref{fig:recrystal_all_tempscalings}.}
	\label{fig:heat_treat_params}
\end{figure*}

We have discussed  how the behavior of an ion trap was affected by ten heat treatments. In Fig.~\ref{fig:heat_treat_params} we present the details of these heat treatments, plotting how the temperature of the ion trap varied as a function of time.

 The ion trap was milled between each heat treatment. In some cases the trap was milled multiple times between these heat treatments, or bombarded with electrons at the ion-heating-rate measurement region. After each milling step or electron treatment, the temperature dependence of the ion heating rates was measured. These measurements were performed more quickly and at lower temperatures than those reached during heat treatments. 
 
The parameters of all argon ion milling surface treatments can be found in Tab.~\ref{tab:mill}. This table describes the milling treatments and electron bombardments in chronological order. This chronology also includes the ten heat treatments to provide the necessary context as to when these treatments were performed. Additional information on the full trap history can be found in Ref~\cite{Berlin-Udi2020TreatmentsSurface}.

\begin{center}
\begin{table}[h!]
  \begin{center}
    \caption{Chronology of argon ion milling, electron bombardment, and heat treatments, with details of milling parameters.}
    \label{tab:mill}
    \begin{tabular}{| l | r | r | r | r | r | r |}
      \hline
      \textbf{Treatment} & \textbf{Energy}         &   \textbf{Beam}    & \textbf{Beam}    \\
                         & \textbf{deposited}      &   \textbf{Angle}   & \textbf{Energy}  \\  
                         & (J/cm$^2$)              &                    & (eV)             \\ 
      \hline  
      \textsc{heat 1} & --- \ \ \ \ \ \ & --- \ \ \ & --- \ \ \ \ \\
      Mill 0.13~nm    & $0.1$                   &   normal           & 200              \\
      \textsc{heat 2} & --- \ \ \ \ \ \ & --- \ \ \ & --- \ \ \ \ \\
      Mill 0.13~nm    & $0.1$                   &   normal           & 200              \\
      \textsc{heat 3} & --- \ \ \ \ \ \ & --- \ \ \ & --- \ \ \ \ \\
      Mill 0.6~nm     & $0.5$                   &   normal           & 200              \\
      \textsc{heat 4} & --- \ \ \ \ \ \ & --- \ \ \ & --- \ \ \ \ \\
      Mill 1.5~nm     & $1.2$                   &   normal           & 200              \\
      Mill 3.9~nm     & $2.9$                   &   normal           & 400              \\
      \textsc{heat 5} & --- \ \ \ \ \ \ & --- \ \ \ & --- \ \ \ \ \\
      Mill 11.4~nm    & $7.5$                   &   normal           & 400              \\
      \textsc{heat 6} & --- \ \ \ \ \ \ & --- \ \ \ & --- \ \ \ \ \\
      Mill 16.4~nm    & $8.8$                   &   normal           & 200              \\
      \textsc{heat 7} & --- \ \ \ \ \ \ & --- \ \ \ & --- \ \ \ \ \\
      Mill 0.6~nm     & $0.2$                   &   normal           & 200              \\
      Mill 1.3~nm     & $0.6$                  &   normal           & 200              \\
      Electron treat  & --- \ \ \ \ \ \ & --- \ \ \ & --- \ \ \ \ \\
      \textsc{heat 8} & --- \ \ \ \ \ \ & --- \ \ \ & --- \ \ \ \ \\
      Mill 12.2~nm    & $4.9$                  &   normal           & 200              \\
      \textsc{heat 9} & --- \ \ \ \ \ \ & --- \ \ \ & --- \ \ \ \ \\
      Electron treat  & --- \ \ \ \ \ \ & --- \ \ \ & --- \ \ \ \ \\
      Mill 18.9~nm    & $9.7$                  &   45$^\circ$       & 400              \\
      Mill 9.8~nm     & $5.2$                  &   45$^\circ$       & 500             \\
      \textsc{heat 10}& --- \ \ \ \ \ \ & --- \ \ \ & --- \ \ \ \ \\
      Electron treat  & --- \ \ \ \ \ \ & --- \ \ \ & --- \ \ \ \ \\
      \hline
      \end{tabular}
      \end{center}

\end{table}
\end{center}

\section{Thermally activated fluctuators}\label{app:taf}

The results of the first ion heating rate measurements performed in this trap were consistent with the thermally activated fluctuator (TAF) model, as discussed in a previous publication \cite{Noel2019-TAF}.

\begin{figure}[ht!]
	\centering
	\begin{subfigure}[t]{.48\textwidth}
		\includegraphics{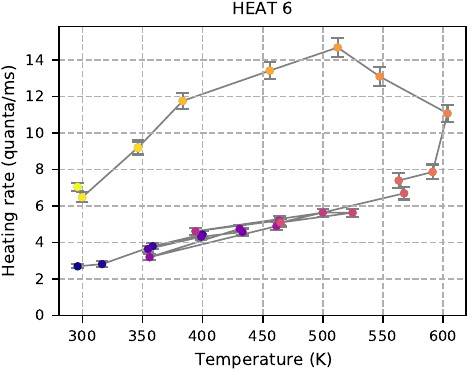} 
	\end{subfigure} \\
	\vspace{.3cm}
	\begin{subfigure}[t]{.48\textwidth}
		\includegraphics{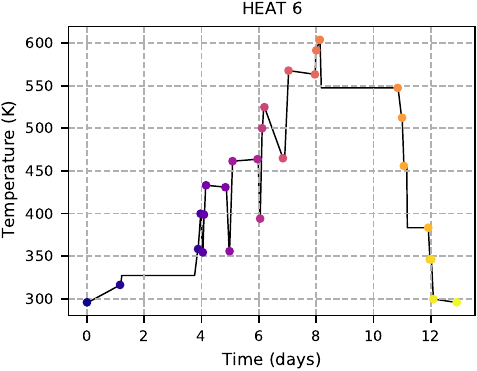} 
	\end{subfigure}%
	\caption{During treatment \textsc{heat 6}, ion heating rates were measured as the trap temperature was raised and lowered in an alternating manner. No thermal transformation took place until after the temperature had been raised above 550~K. These measurements were performed at trap frequency $2\pi \times 0.88$~MHz.}
	\label{fig:lowtemp_notransform}
\end{figure}

Before we use the TAF model to make predictions about the experiments discussed in this manuscript, certain measurements must be excluded that are known to be outside of this model, such as those taken during active thermal transformations. Although noise during an active transformation may still be dominated by TAFs, temperature-scaling slopes and frequency-scaling exponents lose physical significance when each measurement samples a different underlying TAF distribution. We performed an experiment to determine at what temperatures thermal transformations begin to take place. The results of this experiment are shown in Fig.~\ref{fig:lowtemp_notransform}, where data from \textsc{heat 6} are plotted. During the first seven days of this heat treatment, we alternated between raising and lowering the temperature of the substrate between measurements. During this portion of the experiment, ion heating rates increased as the temperature was raised, and they decreased back to previously-measured values when the temperature was lowered. The transformation did not take place until the temperature of the trap was raised above 550~K. With this in mind, we have chosen to  exclude all data taken above 550~K. This also removes the feature in which the heating rates abruptly increase in magnitude, as was shown in Fig.~\ref{fig:full_range_temp_scaling}. Because of its sharp rise, we believe that this feature is due to a distinctly different noise mechanisms than at lower temperatures. This observation further supports the choice to exclude data at temperatures above 550\;K from the analysis.

After high-temperature data is excluded, the data is binned and averaged in 10~K increments. Then, we fit a Gaussian to each heating-rate temperature dependence to obtain noise distributions with smoothly-varying slopes. The choice to use a Gaussian fit is not a statement about the underlying physics of this system; it is simply versatile enough to fit the data of interest. 

In order to use the smoothed heating-rate temperature dependence to make a prediction of how $\alpha$ scales with temperature, a value must be chosen for the attempt time $\tau_0$. In the context of defect dynamics in solid state systems, attempt times between $10^{-12}$ and $10^{-14}$~s have been used, as this is on the order of the inverse phonon frequency \cite{Dutta1981-review, Briggmann1994IrradiationinducedNoise, Koch19851fAlloys}. Attempt times on the same order have been used in the context of adatom diffusion on metal surfaces \cite{Barth2000TransportPrecursors}. Defect fluctuations and adsorbate motion are both viable candidates for the physical manifestation of TAFs in our system, so we perform our analysis using an attempt time of $10^{-13}$~s. The dependence of $\alpha$ on $\tau_0$ is logarithmic, so it is not critical to know the precise value of $\tau_0$.

TAF noise from a slowly-varying distribution of fluctuators scales inversely with frequency, locally, according to:
\begin{equation}
\label{eqn:taf_alpha_1}
S(\omega, T) \propto 1/\omega^{\alpha}
\end{equation}
where $\alpha$ is the frequency scaling exponent, and the measurement frequency $\omega$ is equal to the secular frequency of the ion. The TAF model predicts that $\alpha$ will depend on the slope of the temperature scaling according to \cite{Dutta1979}:

\begin{equation}
\label{eqn:taf_alpha_2}
\alpha (\omega,T) = 1-\frac{1}{\ln (\omega \tau_0)} \left(\frac{\partial \ln S(\omega, T)}{\partial \ln T} -1\right) .
\end{equation}

Where $\tau_0$ is the attempt time. It follows from Eq.~\ref{eqn:taf_alpha_2} that if the slope of the ion heating rate temperature scaling changes with temperature, so does $\alpha$. Thus, we can measure the slope of the temperature scaling of ion heating rates in our system, and use Eq.~\ref{eqn:taf_alpha_2} to predict the temperature scaling behavior of $\alpha$.

To quantify the accuracy of the TAF prediction, we perform a reduced $\chi^2$ analysis comparing the measured values of $\alpha$ to the predictions from the TAF model extracted from the Gaussian fit of the 2$\pi \times 1.3$~MHz heating-rate temperature dependence. The $\chi^2_r$ values are calculated as follows:
 
\begin{equation} 	
	\chi^2_r = \frac{1}{n}\sum_{i=1}^{n} \left(\frac{\alpha_i - \alpha_{\textsc{taf}}(T_i)}{\sigma_i}\right)^2
\end{equation}

where $\alpha_i$ is a direct measurement of the frequency scaling exponent, $\sigma_i$ is the standard error of $\alpha_i$, and $T_i$ is the temperature at which $\alpha_i$ was measured. $\alpha_{\textsc{taf}}(T_i)$ is the value of $\alpha$ at temperature $T_i$ predicted by the TAF model using one of the measured temperature dependencies. $n$ is the total number of $\alpha$ measurements in this data-set.

\section{Exploration of atomic restructuring}\label{ap:restructuring}

While we cannot make a definitive statement on what physical processes drive thermal transformations, we observe that atomic restructuring of the metal appears to be consistent with our experimental results. We now will explore how this restructuring could have physically manifested in our metal film. We are not able to take \textit{in situ} measurements of surface morphology or crystal structure, but we are aided by the observation that ion milling acts in opposition to thermal transformations: thermal noise transformations cause ion heating rates to rise, and milling brings them back down. In addition, after a thermal transformation has saturated, milling can prime the surface for further transformation to take place. We can use the observation that milling and heating act in opposition to develop hypotheses about what physical processes could be causing thermal noise transformations. 
 
Milling disrupts the crystal structure of a thin film by knocking atoms out of their positions in the lattice. Niu {\em et al.}~\cite{Niu2021} bombarded aluminum surfaces with argon ions, and HRTEM cross-section images showed that ion bombardment produced a 400~nm-deep amorphous layer at the surface of previously-crystalline aluminum. Since collisions cascade through the metal, ion bombardment can produce extended layers of structural disorder without removing a large amount of material \cite{Barber1993}. This atomic scale disorder is highly energetically unfavorable compared to a more crystalline structure. Heating counteracts sputter-induced amorphization and restores crystallinity through thermodynamically favorable rearrangements. 
 
Yet another way in which annealing and milling oppose each other is by rearranging impurities in the film. During annealing, the recrystallization process can eject impurities from the bulk of a metal and cause them to cluster at grain boundaries \cite{Tytko2012MicrostructuralTomography,de_hass_grain_2001,michael_grain_1991}.
In contrast, ion milling removes material, drives surface adsorbates into the bulk, and implants incident ions \cite{Thome1994AmorphousMixing}. We did not observe a significant change in contaminant concentration after thermal transformations, which indicates that there was not a large migration of contaminants from the bulk to the surface. However, this does not rule out impurity clustering at grain boundaries. 

We hypothesize that carbon, one of the major impurities detected in our trap, would segregate to the grain boundary of an aluminum trap at 550~K. Carbon has an atomic radius smaller than that of silicon, and it has the same valence electronic structure, so its interactions with an aluminum lattice will likely be similar to those of silicon. Simulations of silicon segregation at the grain boundary sites of aluminum indicates silicon grain boundary enrichment is energetically favorable at most sites \cite{wang_first-principles_2015,karkina_solutegrain_2016}. Grain boundary precipitants rich in Si were observed in low-temperature annealing study of two different silicon-containing aluminum alloys after $\approx$ 20 hours at 430~K \cite{de_hass_grain_2001}, suggesting solutes with similar properties, like carbon, would do the same. 

We now turn our attention to changes in the surface morphology. Annealing can grow micron-scale features such as hillocks and other strain-related structures at the surfaces of metal films \cite{Ri2012-hillock-cycle}. Ion bombardment can be used to mill down such features, as material sputtered from valleys has a higher re-deposition rate than material sputtered from peaks. In our experiments, the removal of 0.6~nm of material via argon ion milling had a significant impact on noise from a thermally transformed substrate, as shown in Fig.~\ref{ch:heat_fig:little_mill}. The subsequent milling removed an additional 1.3~nm of material and reversed nearly the entire effect of the thermal transformation. This is evidence that thermal noise transformations cannot be explained by the growth of large features such as hillocks. Ion bombardment can be used as a tool to polish surfaces and remove micron-scale features, but to do so would require more than 2~nm of material removal. 

Instead we believe that electric field noise is likely caused by mechanisms acting on atomic length scales. Indeed, simulations of adsorbate vibrations show that significant electric-field noise arises from adsorbate motions over nanometer and atomic length scales \cite{Ray2019, Safavi-Naini2013}. Annealing increases grain size and crystallinity at atomic length scales. This behavior has been measured in previous experiments on aluminum \cite{Wibowo2017}. In contrast, experiments have shown that ion milling increases the atomic-scale surface roughness of metals \cite{Valbusa2003NanostructuringBeam, Rusponi1998ScalingCu110}. In addition, milling-induced roughness at the atomic scale may be particularly enhanced at the surface of our film, as this surface contains a variety of elements with different sputter rates. The sputter rate of aluminum bombarded with 200\;eV argon ions, for example, is an order of magnitude greater than that of carbon \cite{Yamamura1996}.

\end{document}